\documentclass[aps,pre,twocolumn,groupedaddress,longbibliography]{revtex4-1}
\usepackage[dvipsnames]{xcolor}
\usepackage{graphicx,amsmath,amssymb,color,soul}

\newcommand{\changedforRone}[1]{\textcolor{red}{#1}}
\newcommand{\changedforRtwo}[1]{\textcolor{blue}{#1}}
\newcommand{\changedforRthree}[1]{\textcolor{cyan}{#1}}
\newcommand{\changedformultiple}[1]{\textcolor{purple}{#1}}
\newcommand{\otherchanges}[1]{\textcolor{ForestGreen}{#1}}

\begin{document}

\title{Ultrastable jammed sphere packings with a wide range of particle dispersities}
\author{Robert S. Hoy}
\email{rshoy@usf.edu}
\affiliation{Department of Physics, University of South Florida, Tampa, Florida 33620,USA}
\date{\today}
\begin{abstract}
We show that for a standard continuously-polydisperse model with particle-diameter distribution $P(\sigma) \propto \sigma^{-3}$ and polydispersity index $\Delta$, employing a combination of standard SWAP moves and transient degree of freedom (TDOF) moves during a Lubachevsky-Stillinger-like particle-growth process dramatically increases the generated packings' jamming densities $\phi_{\rm J}(\Delta)$ and coordination numbers $Z_{\rm J}(\Delta)$, for a wide range of $\Delta$.
The fractional increase in $\phi_{\rm J}(\Delta)$ obtained by employing these moves first increases rapidly with $\Delta$, then plateaus at $6-7\%$ over the range $0.10 \lesssim \Delta \leq 50$; the obtained $\phi_{\rm J}$ are as high as $0.747$ (for $\Delta = 0.50$).
These density increases are achieved without producing \changedforRtwo{clearly-noticeable} crystallization or \changedforRtwo{increased} fractionation\changedforRone{, despite the fact that the packings are quite hyperstatic for all $\Delta > 0$.}
SWAP and TDOF moves also reduce packings' rattler populations by as much as \changedforRtwo{99\%} and increase their bulk moduli by as much as \changedforRtwo{80\%}.
\end{abstract}
\maketitle

\section{Introduction}
\label{sec:intro}

Studies of the jamming transition's preparation-protocol dependence arguably began with Scott's observation that steel ball bearings poured into a sufficiently large container tended to form ``random loose packed'' solids with packing fractions $\phi = \phi_{\rm RLP} \simeq 0.59$ \cite{scott60}.
Under ``tapping'' (mechanical excitation), these systems settled further, eventually forming ``random close packed'' solids with packing fractions $\phi = \phi_{\rm RCP} \simeq 0.64$ \cite{scott60,scott69,knight95}.
Removing the mechanical excitation at any point during such a tapping experiment produces a jammed state with $\phi_{\rm RLP} < \phi_{\rm J} < \phi_{\rm RCP}$ \cite{knight95}.
This range ($\phi_{\rm RLP} < \phi_{\rm J} < \phi_{\rm RCP}$) is specific to low-friction, non-cohesive, low-dispersity spheres that were initially rapidly poured into a container.
Increasing the strength of frictional and/or cohesive interactions, changing the particle size distribution, and changing the initial preparation protocol can all produce different ranges for $\phi_{\rm J}$ \cite{carson98,blum04,castellanos05,parteli14,liu15,liu17,nan23}.

Issues like these led Torquato \textit{et al.}\ to ask the question ``Is random close packing well-defined?'', and demonstrate using simulations that it is not \cite{torquato00}.
They found that the varying the particle-growth rate $\Gamma$ in the Lubachevsky-Stillinger (LS) algorithm \cite{lubachevsky91} produces monodisperse frictionless-hard-sphere (MFHS) packings with a wide range of jamming densities $0.64 \lesssim \phi_{\rm J} \lesssim 0.68$, with larger $\Gamma$ producing lower $\phi_{\rm J}$.
Using these results, they proposed that jammed packings exist over a finite domain in ($\phi_{\rm J},\Psi$)-space, where $\Psi$ is some (as yet undefined) scalar \changedforRtwo{structural-}order parameter, and that the traditional estimate of $\phi_{\rm RCP}$ (0.637 \cite{scott60}) actually corresponds to the ``MRJ'' packing fraction $\phi_{\rm MRJ}$, i.e, the packing fraction of \textit{maximally random} jammed states that \textit{minimize} $\Psi$ \cite{torquato00,donev04c}.
In simulations of MFHS, $\phi_{\rm MRJ}$ is the jamming density obtained by starting from a state with a packing fraction below the ``onset'' density $\phi_{\rm on} \simeq 0.45$ and then rapidly compressing it.

Since then, hundreds of studies have examined the structure of amorphous sphere packings with $\phi_{\rm J} \simeq \phi_{\rm MRJ}$.
Denser packings have received much less attention, but the preparation-protocol dependence of $\phi_{\rm J}$ and $\Psi$ has attracted substantial interest.
In particular, several studies \cite{chaudhuri10,morse21,ozawa17} have examined onset effects in hard sphere liquids that were first thermally equilibrated at various packing fractions $\phi_{\rm eq}$ and then jammed using  ``crunching'' protocols that apply compression in the limit $\Gamma \to \infty$.
These studies found that both $\phi_{\rm J}$ and various metrics quantifying $\Psi$ increase with increasing $\phi_{\rm eq}$, provided $\phi_{\rm eq} > \phi_{\rm on}$.

More specifically, Chaudhuri \textit{et al.}\ found that the standard 50:50 1:1.4 bidisperse hard sphere model's $\phi_{\rm J}$ increased from $0.648$ to $0.662$ as $\phi_{\rm eq}$ increased from $0.357$ to $0.584$ \cite{chaudhuri10}.
Charbonneau and Morse found that results for $\phi_{\rm J}(\phi_{\rm eq})$ can be fit by a simple functional form, and also showed that increasing $\phi_{\rm J}(\phi_{\rm eq})$ are also found in higher spatial dimensions \cite{morse21} despite the fact that they are absent in mean-field theories of the glass/jamming transition.
Finally, Ozawa \textit{et al.}\ exploited the SWAP Monte Carlo algorithm's ability to equilibrate polydisperse HS liquids at larger $\phi_{\rm eq}$ \cite{berthier16b} to obtain much higher $\phi_{\rm J}$ \cite{ozawa17}.
For a continuously-polydisperse particle size distribution with $P(\sigma) \propto \sigma^{-3}$ for $\sigma_{\rm min} \leq \sigma \leq \sigma_{\rm max}$ and polydispersity index $\Delta = \langle \sigma^2 \rangle/\langle \sigma \rangle^2 - 1 = 0.23$, they found $\phi_{\rm J}$ as high as $0.698$.
These trends arise because liquids prepared at $\phi_{\rm eq} < \phi_{\rm on}$ fall out of equilibrium at $\phi_{\rm on}$ under rapid compression, but liquids prepared at $\phi_{\rm eq} > \phi_{\rm on}$ necessarily fall out of equilibrium at higher densities $\phi_{\rm g}$, and their  greater structural order at these $\phi_{\rm g}$ ultimately 
causes them to have larger $\phi_{\rm J}$ and $\Psi$ \otherchanges{after crunching.}

Such results have deepened our understanding of what jamming entails, and are of broad interest given that analogous onset phenomena are observed in \textit{thermal} systems.
Glass-forming liquids equilibrated  at fixed $\phi$ and temperatures $T$ above the onset temperature $T_{\rm on}(\phi)$ always have the same average inherent structure energy ($E_{\rm IS}$), while those equilibrated at temperatures $T <  T_{\rm on}$ have $E_{\rm IS}$ that decrease with decreasing $T$ \cite{sastry98,debenedetti01}.
Moreover, if one defines $\phi_{\rm J}$ as the packing fraction below which the average inherent structure energy  $E_{\rm IS} = 0$, then purely-repulsive soft-sphere glasses have $\phi_{\rm J}(t_{\rm eq})$ that increase with the equilibration/aging time $t_{\rm eq}$ \cite{interiano24}.

How far can we take such effects?
Refs.\ \cite{chaudhuri10,morse21,ozawa17} all employed athermal, crunching-like compression algorithms which operated in the limit $\Gamma \to \infty$.
Can one obtain \textit{amorphous} jammed packings with even larger $\phi_{\rm J}$ and $\Psi$ by designing a compression algorithm which \textit{effectively} operates in the limit $\Gamma \to 0$?
Berthier \textit{et al.}\ recently  showed that the answer is ``yes'' \cite{ghimenti24,berthier24}.
By employing an efficient event-driven Monte Carlo routine for the dynamics, and performing ``cluster'' SWAP (cSWAP) moves during both the initial equilibration at $\phi = \phi_{\rm eq}$ and the subsequent compression, they achieved $\phi_{\rm J}$ as large as $0.720$ for the abovementioned $P(\sigma) \propto \sigma^{-3}$ model with $\Delta = 0.23$.
Comparable increases in packing efficiency have been achieved for disks, using various approaches \cite{hagh22,bolton24,kim24}.
In all cases, they are achieved by employing ``unphysical'' moves which speed up systems' dynamics.

Such increases are not specific to disks and spheres.
We recently showed that  they also occur in systems of two-dimensional ellipses \cite{hoy24b}.
By combining a  LS-like particle-growth process with (i) \textit{biased} SWAP Monte Carlo (BSMC) moves which swap the diameters of larger particles with smaller interparticle gaps with those of smaller particles with larger interparticle gaps, and (ii) transient degree of freedom (TDOF) moves which allow particles to grow by different amounts \cite{hagh22,bolton24}, we generated amorphous packings with substantially higher ($\phi_{\rm J},Z_{\rm J}$) than were obtained in any previous studies \cite{delaney05,donev07,vanderwerf18,rocks23}, for a wide range of aspect ratios $\alpha$.
Fractional increases in both $\phi_{\rm J}$ and $Z_{\rm J}$ were strongly and nontrivially $\alpha$-dependent.
The former decreased from $\simeq 5\%$ to $\simeq 1\%$ over the range $1\leq \alpha \lesssim 1.6$, then increased from $\simeq 1\%$ to $\simeq 7\%$ over the range $1.6 \lesssim \alpha \leq 5$.
The latter were also nonmonotonic in $\alpha$.
In particular, the intermediate-$\alpha$ packings were (in contrast to those found in all previous studies \cite{delaney05,donev07,vanderwerf18,rocks23}) \textit{isostatic}, while higher-$\alpha$ packings actually had \textit{lower} $Z_{\rm J}$ because they included  increasingly-large locally-nematic domains reminiscent of liquid glasses \cite{roller20,roller21}.

A natural followup question that remains unanswered is:\ does the protocol-dependence of  polydisperse spheres'  ($\phi_{\rm J},\Psi$) couple comparably strongly and nontrivially to $\Delta$?
In this paper, we show that the answer to this question is ``yes.''
Employing standard SWAP and TDOF moves during a LS-like particle-growth process increases $\phi_{\rm J}$ by at least  6\% for all $0.1 \lesssim \Delta \leq 0.5$.
The increases in $Z_{\rm J}$ are even more dramatic, and are strongly $\Delta$-dependent.
In particular, as $\Delta$ increases \changedforRtwo{beyond $0.25$}, employing TDOF moves at the end of the particle-growth process increasingly allows small rattlers which would otherwise not be part of the packings' percolating, mechanically-rigid ``backbones'' \cite{moukarzel98} to continue growing \otherchanges{until} they too jam, ultimately producing structures that are nearly rattler-free and very mechanically stable.
For example, this procedure yields fractional density increases, rattler-population reductions, and bulk-modulus increases of 6.5\%, \changedforRtwo{96\%}, and \changedforRtwo{80\%} for $\Delta = 0.50$ \changedforRtwo{without producing clearly-noticeable changes in commonly-investigated $\Psi$-metrics such as the pair correlation function $g(r)$ or the static structure factor $S(q)$.}

\changedforRtwo{The outline of the remainder of this paper is as follows.
Section \ref{sec:methods} describes our computational model and methods.
Section \ref{sec:results} highlights several key differences between the structural and mechanical properties of packings generated with and without SWAP and TDOF moves, and analyzes how the strength and character of these differences evolves with increasing $\Delta$.
In Section \ref{sec:conclude}, we discuss these trends and their implications for future work.
Finally, Appendices A and B provide additional details.}

\section{Methods}
\label{sec:methods}

We employ \otherchanges{a power-law} particle size distribution 
\begin{equation}
P_\mathcal{R}(\sigma) = \Bigg{ \{ } \begin{array}{ccl}
\displaystyle\frac{2}{\mathcal{R}-\mathcal{R}^{-1}}  \sigma^{-3} & \ \ , \ \ & \sigma_{\rm min} \leq \sigma \leq \mathcal{R}\sigma_{\rm min} \\
\\
0 & \ \ , \ \ & \sigma < \sigma_{\rm min}  \ \rm{or}\ \sigma > \mathcal{R}\sigma_{\rm min} 
\end{array}.
\label{eq:Pofsigma}
\end{equation}
\otherchanges{that has been used in many recent studies of both} glassy dynamics and jamming \otherchanges{\cite{ozawa17,berthier24,interiano24,anzivino23,monti22,monti23,srivastava25,oquendo20,oquendo22,berthier16,ninarello17,scalliet22}}.
Here $\mathcal{R} \geq 1$ is the ratio of the diameters of the largest and smallest particles.
This distribution gives \otherchanges{$\langle\sigma\rangle = 2\mathcal{R}\sigma_{\rm min}/(\mathcal{R}+1)$, $\langle\sigma^2\rangle = 2\mathcal{R}^2\ln(\mathcal{R} )\sigma_{\rm min}^2/(\mathcal{R}^2 -1)$}, and
\begin{equation}
\Delta(\mathcal{R}) \equiv \displaystyle\frac{\sqrt{\langle\sigma^2\rangle - \langle\sigma\rangle^2}}{\langle\sigma\rangle} =  \sqrt{\displaystyle\frac{\mathcal{R} +1}{2(\mathcal{R} -1)} \ln(\mathcal{R}) - 1}.
\label{eq:deltaform}
\end{equation}
Most previous studies employing this $P(\sigma)$ employed \otherchanges{$\Delta \leq 0.23$}.
\otherchanges{Ref.\ \cite{anzivino23} examined both a closely-spaced set of $0 \leq \Delta \lesssim 0.65$ (a range slightly wider than that considered here) and selected larger $\Delta$.}
\otherchanges{Refs.\ \cite{oquendo20,oquendo22,monti22,monti23,srivastava25} examined selected $0.20 \leq \Delta \leq 1.16$}, and found $\phi_{\rm J}$ as high as  \otherchanges{$\simeq 0.82$} in very large systems with $\Delta \gtrsim 1$.
However, \otherchanges{none of these studies employed} unphysical moves during sample compression, suggesting that even higher jamming densities can be obtained when such moves are employed.

Here, to systematically characterize how employing SWAP and TDOF moves during particle growth affects ($\phi_{\rm J},\Psi$), we contrast packings generated with and without these moves  for a wide range of narrowly spaced $\Delta$; specifically, $\Delta = \changedforRtwo{0.01}i\ \forall\ i \in \{0,1,2,...,50\}$.
We begin by placing $N$ hard spheres with the corresponding $P_\mathcal{R}(\sigma)$ randomly within cubic periodic cells while forbidding interparticle overlaps.
The initial \otherchanges{$\sigma_{\rm min}$} is chosen to be small enough that the initial packing fraction $\phi_{\rm init} < 10^{-3}$.
Then we employ a LS-like particle growth algorithm consisting of two stages per cycle:
\begin{enumerate}
\item Attempting to translate each particle $i$ by a random displacement along each Cartesian direction; and
\item Increasing \textit{all} particles' diameters $\sigma$ by the \textit{same} factor $\min(\tilde{\mathcal{G}}/10,\Gamma_{\rm max})$, where $\tilde{\mathcal{G}}$ is the value that brings \textit{one} pair of spheres into tangential contact, and $\Gamma_{\rm max} = 10^{-5}$ is the maximum growth rate.
\end{enumerate}
Here $\tilde{\mathcal{G}} = \min(\mathcal{G}_i)$, where 
\begin{equation}
\mathcal{G}_i = \min\left[ \displaystyle\frac{r_{ij}}{\sigma_{ij}} -1 \right],
\label{eq:fracgrowthrate}
\end{equation}
$r_{ij}$ is the  center-to-center distance between particles $i$ and $j$, and $\sigma_{ij} \equiv (\sigma_i + \sigma_j)/2$ is their contact distance if their diameters are $\sigma_i$ and $\sigma_j$.
The minimum in Eq.\ \ref{eq:fracgrowthrate} is taken over all neighbors of particle $i$, and the subsequent minimum defining $\tilde{\mathcal{G}}$ is taken over all $i$.
These choices make the algorithm more efficient by allowing particles to grow faster when the minimal interparticle gaps are larger.
Note that imposing the uniform growth rate $\tilde{\mathcal{G}}$ preserves the \textit{shape} of the particle-size distribution $P_\mathcal{R}(\sigma)$ defined in Eq.\ \ref{eq:Pofsigma}.
In other words, the ratio $\sigma_{\rm max}/\sigma_{\rm min} = \otherchanges{\mathcal{R}}$ of the largest and smallest particle diameters, and indeed the ratios of all other moments of $P(\sigma)$, remain constant as $\langle \sigma \rangle = \int_{\sigma_{\rm min}}^{\sigma_{\rm max}} \sigma P(\sigma) d\sigma$ increases.

In runs employing SWAP, a third stage is added to each growth cycle:
\begin{enumerate}
\setcounter{enumi}{2}
\item $N$ moves attempting to exchange the diameter of a randomly chosen particle $i$ with that of another randomly chosen particle $j$.  Moves are accepted if they do not produce any interparticle overlaps.
\end{enumerate}
This is just the standard hard-particle SWAP algorithm \cite{grigera01} with an attempt rate of 100\%, i.e., an average of one SWAP attempt per particle per cycle.

As in Ref.\ \cite{hoy24b}, the attempted translations during stage (1) have maximum magnitude $0.05f$, where the move-size factor $f$ is set to $1$ at the beginning of all runs, and is multiplied by $3/4$ whenever $100$ consecutive growth cycles have passed with $\tilde{\mathcal{G}} < 10^{-10}$.
The LS-like algorithm terminates when $f$ drops below $2\times 10^{-9}$, which is the smallest value allowed by our double-precision numerical implementation.
In the runs that do not employ \otherchanges{TDOF moves}, systems are considered jammed at this point.

In those that do, we reset $f$ to \changedforRtwo{$0.01$}, and then execute a second LS-like growth process that operates like the one described above, with one critical difference: each particle is grown by a \textit{different} factor, specifically $\mathcal{G}_i/2$, during each cycle.
This procedure allows the \textit{shape} of $P(\sigma)$ to vary.
These ``transient degree of freedom'' (TDOF) moves\otherchanges{, which are similar in sprit to those performed in Refs.\ \cite{hagh22,bolton24},} continue until all $N$ particles have $\mathcal{G}_i = 0$, i.e.\ until no particles can grow further without producing an overlap.
Systems are then considered jammed.

We assess the mechanical stability of these jammed packings by measuring their bulk moduli $K = \phi \frac{\partial P}{\partial \phi}$, where $P$ is pressure.
To obtain finite $P$ and $K$, we switch from a hard-sphere to a Hertzian pair potential
\begin{equation}
U(r) =  \Bigg{ \{ } \begin{array}{ccl} 
\displaystyle\frac{2\varepsilon}{5}  R_{ij}^{1/2} (\sigma_{ij} - r)^{5/2}  & \ \ , \ \ & r \leq \sigma_{ij}, \\
\\
0 & \ \ , \ \ & r > \sigma_{ij},
\end{array}
\label{eq:hertzpot}
\end{equation}
where $\varepsilon$ is an arbitrary energy scale and $R_{ij} = \sigma_i \sigma_j/(4\sigma_{ij})$ is the reduced radius of particles $i$ and $j$.
We chose this $U(r)$ because it models interparticle contact more realistically than the more commonly employed harmonic potential \cite{ohern03}, e.g., $U(r) \propto R_{ij}^{1/2}$ captures the grain-size dependence of the elastic-strain energy.

We set particles' mass density to $6/\pi$, so they have mass $m_i = \sigma_i^3$.
Then we perform molecular dynamics (MD) simulations that compress systems at a true volumetric strain rate $\dot{\epsilon}_{\rm V} = \dot{V}/V = -10^{-6}/\tau$ until the true volumetric strain $\epsilon_{\rm V} = \ln( \frac{V}{V_{\rm init}}) = -0.1$.
Here $\tau = \sqrt{\langle m \sigma^2 \rangle/\varepsilon}$ is the MD time unit, and the timestep $\delta t = 0.005\tau$.
Simulations are performed using \texttt{hdMD} \cite{hoy22}.

All results presented below are averaged over at least 25 independently-prepared samples.

\section{Results}
\label{sec:results}

\subsection{Jamming densities and coordination numbers}
\label{subsec:IIIA}

\begin{figure}[htbp]
\includegraphics[width=3in]{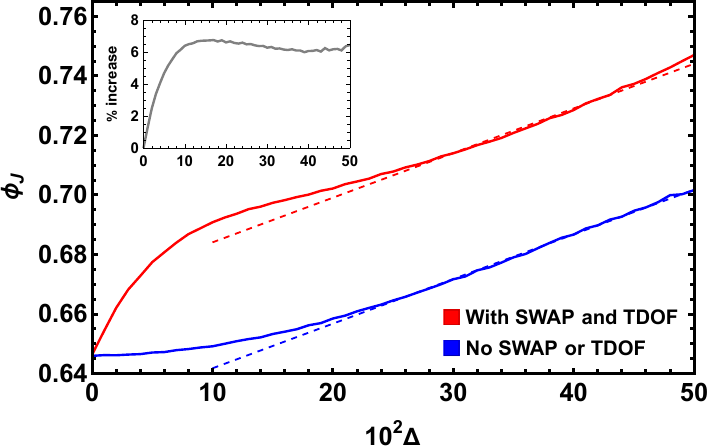}
\caption{Protocol dependence of jamming densities for $N = 1000$.  The inset shows the fractional increase in $\phi_{\rm J}$ obtained by employing SWAP and TDOF moves.   Dashed lines show linear fits to Eq.\ \ref{eq:phiform1} for $\Delta \geq 0.25$.}
\label{fig:1}
\end{figure}

Figure \ref{fig:1} shows the average jamming densities obtained with and without SWAP and TDOF moves for $N = 1000$, which is the same system size employed in Ref.\ \cite{berthier24}.
Without these moves, small polydispersities ($\Delta \leq \changedforRtwo{0}.05$) produce a nearly constant $\phi_{\rm J} \simeq 0.646$, a typical value for low-dispersity frictionless spheres' $\phi_{\rm RCP}$ \cite{ciamarra10b}.
Over the range $0.05 \lesssim \Delta \lesssim 0.25$, both $\phi_{\rm J}$ and $\partial \phi_{\rm J}/\partial \Delta$ gradually increase.
Finally, for $\Delta \gtrsim 0.25$, $\partial \phi_{\rm J}/\partial \Delta$ remains roughly constant, and $\phi_{\rm J}(\Delta)$ increases steadily to a maximum of $\simeq 0.702$ for $\Delta = 0.50$.
Reasons for the increase in $\phi_{\rm J}$ with $\Delta$ have been discussed at length in Refs.\ \otherchanges{\cite{desmond14,oquendo20,oquendo22,monti22,monti23,anzivino23,srivastava25}}.
A simplistic version of the main reason is that increasing $\Delta$ allows small particles to fill in the gaps between their larger neighbors.
This process, of course, does not become possible until $\Delta$ is sufficiently large.

Our results are entirely consistent with this qualitative picture.
\changedformultiple{Remarkably, they are predicted to within less than $ 0.3\%$ by the formula given in Eq.\ 25 of Ref.\ \cite{anzivino23},}
\begin{equation}
\changedformultiple{\phi_{\rm J}(\Delta) =  \phi_{\rm J}(0) + \displaystyle\frac{a_1 \Delta^2 + a_2 \Delta^4 + a_3 \Delta^6}{1 + b_1 \Delta^2 + b_2 \Delta^4 + b_3 \Delta^6},}
\label{eq:zacconephiJ}
\end{equation}
\changedformultiple{for all $\Delta$.
Here the coefficients $\{a_1,a_2,a_3,b_1,b_2,b_3\}$ represent terms that are calculated using liquid-state theory and are each functions of the monodisperse jamming density $\phi_{\rm J}(0)$ and a single fitting parameter $C_0$.
Further details of this agreement are given in Appendix A.}

\otherchanges{Moreover,} our results for $\Delta \gtrsim 0.25$ are well fit by
\begin{equation}
\phi_{\rm J}(\Delta) = \phi_{\rm 0} + 0.15 \Delta,
\label{eq:phiform1}
\end{equation}
with $\phi_0 \simeq 0.627$.
We believe that this formula should be valid for all $\Delta \lesssim 1$ for systems with this $P(\sigma)$ that were prepared using comparable protocols.
For example, Eq.\ \ref{eq:phiform1} predicts $\phi_{\rm J}(1.16) \simeq 0.80$, while Ref.\ \cite{monti22} found $\phi_{\rm J}(1.16) \simeq 0.81$ in systems with $N = 2.3\times 10^5$.
Since systems' jamming densities tend to scale as $\phi_{\rm J}(N) = \phi_{J}(\infty) - bN^{-\nu}$, where $\nu > 0$ is a protocol-dependent power law \cite{scott60,ozawa17}, the difference between these values (i.e., $0.80$ and $0.81$) could arise from finite-system-size effects.

Employing SWAP during sample compression (or equivalently, during particle growth) increases systems' $\phi_{\rm J}$ because it lowers their $\phi$-dependent $\alpha$ relaxation times $\tau_\alpha(\phi)$ by orders of magnitude \cite{berthier16b,ninarello17}.
This allows them to sample their energy landscapes more efficiently, and hence to avoid the kinetic traps which otherwise lead to jamming \cite{ghimenti24,berthier24}.
Such reductions in $\tau_\alpha(\phi)$ are, of course, absent in monodisperse systems where swapping particles' diameters has no effect.
This implies that they must strengthen with increasing $\Delta$ for ``small'' $\Delta$.
For the algorithm employed here, in which SWAP attempt partners are chosen completely randomly, they must also weaken with increasing $\Delta$ for ``large'' $\Delta$ as swap-move success rates decrease  \cite{berthier16b,ninarello17}.
However,  the details of how such $\Delta$-dependent reductions of $\tau_\alpha(\phi)$ affect the concomitant increases in $\phi_{\rm J}(\Delta)$ have not been previously explored.

As illustrated in Fig,\ \ref{fig:1}, the fractional increases in $\phi_{\rm J}$ enabled by employing SWAP during particle growth initially grow rapidly with increasing $\Delta$, as expected.
We find that the peak packing-efficiency gain ($\simeq 6.7\%$) is achieved over the range $0.13 \lesssim \Delta \lesssim 0.18$.   
As $\Delta$ continues to increase, the gains decrease slightly, reaching a minimum of $\simeq 6.0\%$ (at $\Delta = 0.39$) before gradually increasing again.
The overall effect of these $\Delta$-dependent efficiency gains is to \textit{qualitatively} change the shape of the $\phi_{\rm J}(\Delta)$ curve.
\otherchanges{In particular, it becomes wholly inconsistent with Eq.\ \ref{eq:zacconephiJ}.}
Instead of being concave up for all $\Delta \lesssim 0.25$, it is concave down.
Then,  over the range $0.25 \lesssim \Delta \lesssim 0.45$, it remains linear.
Over this range,  $\phi_{\rm J}(\Delta)$ is again fit by Eq.\ \ref{eq:phiform1}, but now with $\phi_0 \simeq 0.669$.
In other words, employing SWAP and TDOF moves enables a nearly constant packing-fraction increase $\Delta\phi = 0.042$ over this range.

For the well-studied case of $\Delta = 0.23$, we find $\phi_{\rm J} \simeq 0.662$ ($\phi_{\rm J} \simeq 0.705$) for runs that do not (do) employ SWAP and TDOF moves.
The former result is consistent with previous studies employing traditional particle-growth algorithms, low $\phi_{\rm eq}$ and moderate $\Gamma$ \cite{berthier16b,ozawa17}.
The latter result is consistent with the $\phi_{\rm J}$ obtained using Berthier  \textit{et al.}'s more-sophisticated growth algorithm with low $\phi_{\rm eq}$ \cite{berthier24}; the\otherchanges{ir} higher maximum $\phi_{\rm J} \simeq 0.720$ (discussed above) was obtained for $\phi_{\rm eq} \simeq 0.66$.

\changedforRthree{The vast majority of the increases in $\phi_{\rm J}(\Delta)$ discussed above arise from the SWAP moves rather than the TDOF moves.
We have verified that this is so by performing independent runs (not shown in Fig.\ 1) that employed \textit{only} SWAP moves.
The additional fractional increases in $\phi_{\rm J}$ obtained during the second stage of LS-like growth (Section \ref{sec:methods}) decrease from 0.07\% for $\Delta = 0$ to $< 0.01\%$ over the range $0.02 \leq \Delta < 0.30$.
For larger $\Delta$, they increase in a concave-up fashion, reaching 0.05\% by $\Delta = 0.39$ and 0.3\% by $\Delta = 0.50$.
Although these increases are small compared to those highlighted in Fig.\ \ref{fig:1}'s inset, they explain two trends shown in this figure that were not discussed above:}
\changedforRthree{(i)} while the overall packing-efficiency gains  decrease over the range $0.18 \lesssim \Delta \lesssim 0.39$ as SWAP moves become less efficient  (as expected  \cite{berthier16b,ninarello17}), they increase again over the range $0.39 \lesssim \Delta \lesssim 0.50$\changedforRthree{; and (ii)} $\phi_{\rm J}(\Delta)$ increases \changedforRthree{noticeably} above its linear-fit value for the largest $\Delta \gtrsim 0.45$.

\begin{figure}[h!]
\includegraphics[width=3in]{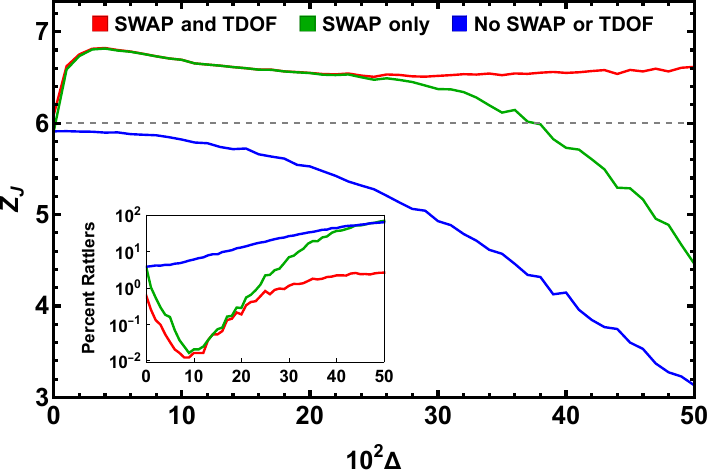}
\caption{Protocol dependence of average coordination numbers and rattler fractions for $N = 1000$ \changedforRtwo{and $C = 10^{-3}$.} The horizontal dashed line indicates $Z_{\rm iso} \equiv 6$.}
\label{fig:2}
\end{figure}

To better understand the above results, we examined the jammed packings' coordination-number statistics.
\changedforRtwo{Figure \ref{fig:2}} shows the average coordination numbers $Z_{\rm J} = N^{-1}\sum_{i = 1}^N Z_i$, where particle $i$ has $Z_i$ contacts.
In contrast to most previous studies, we do \textit{not} remove rattlers (particles with $Z_i < 4$ \cite{noncohem}) when calculating $Z_{\rm J}$, because doing so obscures the trends we wish to highlight.
Here we define contact using the criterion \changedforRtwo{employed in Ref.\ \cite{ozawa17}, i.e.\ $r_{ij} < (1+C) \sigma_{ij}$, with $C = 10^{-3}$.}
\otherchanges{We chose this value because it lies within the well-known intermediate-$C$ plateau in $Z_{\rm J}(\Delta, C)$ \cite{donev05c}.
Reducing the value of $C$ shifts the $Z_{\rm J}(\Delta)$ curves downwards and increases} the rattler fractions, \otherchanges{but} it does not qualitatively change the results discussed below.
\changedforRone{Results for the standard definition of $Z_{\rm J}$ (i.e.\ the average coordination number \textit{after} removing  rattlers) and a further discussion of $C$-dependence are given in Appendix B.}

In systems prepared without SWAP and TDOF moves, $Z_{\rm J} = \changedforRone{5.91}$ for $\Delta = 0$, and decreases monotonically with increasing $\Delta$, to a minimum of $Z_{\rm J} =  \changedforRone{3.13}$ for $\Delta = 0.50$.
Over the same range of $\Delta$, the rattler fraction increases monotonically from  \changedforRone{3.8\%} to  \changedforRone{63\%}.
Both of these trends arise because (as discussed in Section \ref{sec:methods}) our LS-like growth algorithm terminates and systems are considered jammed when $\tilde{\mathcal{G}}$ drops to zero within numerical precision, i.e.\ when the particles within the percolating, mechanically-rigid backbones whose formation normally coincides with jamming \cite{moukarzel98} can no longer grow without producing overlaps.
As $\Delta$ increases, these backbones are increasingly dominated by the large particles \cite{monti22,monti23}, and thus the jammed states contain an increasing number of small rattlers.
We emphasize, however, that these results are \textit{not} artifacts of employing a LS-like particle-growth algorithm; the same trends are present in jammed Hookean-sphere packings prepared via compression at constant pressure  \cite{monti22,monti23}.

\changedforRthree{Adding SWAP moves to the packing-generation procedure yields far-higher $Z_{\rm J}(\Delta)$ for all $\Delta > 0$.
As $\Delta$ increases, $Z_{\rm J}$ first increases rapidly to a maximum of \changedforRone{$6.81$} at  $\Delta = 0.04$, then decreases in a concave-up fashion over the range $0.04 < \Delta \lesssim 0.25$, then decreases in a concave-down fashion, reaching its minimum value \changedforRone{($4.46$)} at $\Delta = 0.50$.
For small polydispersities, these increases in $Z_{\rm J}$ are directly associated with dramatic reductions in the rattler fractions; we find \changedforRone{$f_{\rm ratt} < 0.01$} over the range \changedforRone{$0.02 \leq \Delta \leq 0.23$}.
On the other hand, SWAP moves rapidly lose their propensity to eliminate rattlers as $\Delta$ increases beyond $0.25$, and we find that employing them \textit{increases} the rattler fractions for $\Delta > 0.47$.
Note that $f_{\rm ratt}$ \textit{usually} increases with increasing $\phi_{\rm J}$ for fixed $\Delta$ \cite{ozawa17}.}

Systems prepared with SWAP \textit{and} TDOF moves have far more contacts for both $\Delta = 0$ and all \changedforRone{$\Delta \gtrsim 0.25$}.
For $\Delta = 0$, the TDOF moves increase $\Delta$ from \changedforRone{$5.91$} to \changedforRone{$6.09$}.
With increasing polydispersity, $Z_{\rm J}$ first passes through a local maximum ($Z_{\rm J} \simeq \changedforRone{6.82}$) at $\Delta = \changedforRone{0.04}$, then decreases slowly before plateauing at \changedforRone{$6.52-6.54$} over the range \changedforRone{$0.21 \lesssim \Delta \lesssim 0.37$}.
Finally, for \changedforRone{$\Delta > 0.37$}, $Z_{\rm J}$ increases steadily, to a maximum of \changedforRone{$6.62$} at $\Delta = 0.50$.
As illustrated in the inset, \changedforRthree{adding the TDOF moves produces a further substantial reduction in $f_{\rm ratt}$ for both small and large polydispersity.
The latter trend corresponds directly to the divergence at $\Delta \simeq 0.25$ of the $Z_{\rm J}(\Delta)$ curves for systems generated using both SWAP and TDOF moves and systems generated using only SWAP moves.
We will argue below that this occurs because the TDOF moves allow smaller rattlers present in the system at the end of the first LS-like particle-growth stage (Section \ref{sec:methods}) to continue growing until they also jam.}

\begin{figure}[h!]
\includegraphics[width=3in]{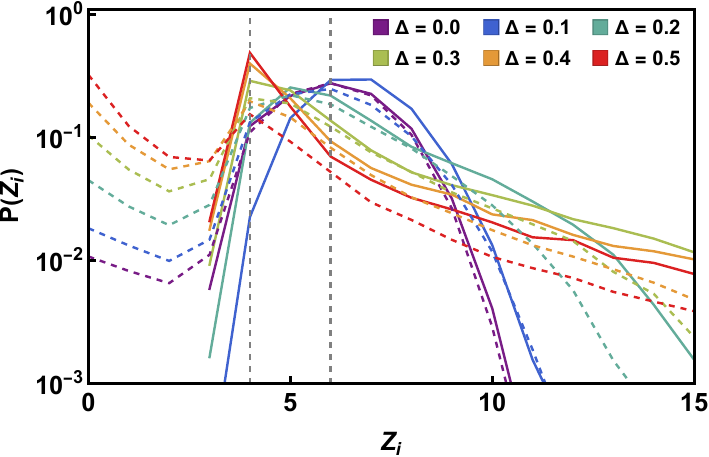}
\caption{Protocol dependence of coordination-number probability distributions \changedforRthree{$P(Z_i)$} for $N = 1000$ \changedforRtwo{and $C = 10^{-3}$.}
Solid and dashed curves show results obtained with and without SWAP and TDOF moves, while the vertical dashed lines indicate $Z_i = 4$ and $Z_i = 6$.  
Increasing/decreasing \changedforRthree{$C$ lengthens/shortens the tails of these distributions, but does not change the qualitative trends shown here.}}
\label{fig:oldfig2b}
\end{figure}

\changedforRthree{The origin of $Z_{\rm J}(\Delta)$'s very-strong protocol-dependence is clarified by examining}  the probability distributions $P(Z_i)$, i.e.\ the probabilities that a given particle has $Z_i$ contacts.
\changedforRthree{Results are shown in Figure \ref{fig:oldfig2b}.} 
Systems prepared without SWAP and TDOF moves have many ``floaters'' (particles with $Z_i = 0$), and also substantial populations of rattlers with contacts (particles with $1 \leq Z_i \leq 3$).
Their $P(Z_i)$ increase rapidly with increasing $\Delta$ for all \changedforRone{$0 \leq Z_i \leq 3$}; \otherchanges{similar results were reported in Refs.\ \cite{oquendo20,oquendo22}.}
Minimally-mechanically-stable particles with $Z_i = 4 \equiv d+1$ (where $d$ is the spatial dimension \cite{noncohem}) exhibit a qualitatively-different trend; $P(4)$ is nonomonotonic in $\Delta$, peaking at $\Delta \simeq \changedforRone{0.34}$.

The \textit{shape} of the $P(Z_i)$ distributions also changes qualitatively as polydispersity increases.
For all $\Delta \leq 0.13$, $P(6) > P(5) > P(4)$.
\changedforRone{Then, for $0.14 \leq \Delta \leq 0.20$, $P(5) > P(6) > P(4)$.
Next, for $0.21 \leq \Delta \leq 0.26$, $P(5) > P(4) > P(6)$.
Finally, for all $\Delta > 0.26$, $P(4) > P(5) > P(6)$.
This smooth} crossover from dominance of locally-isostatic particles with $Z_i = 6$ to minimally-mechanically-stable particles with  $Z_i = 4$ has not (to the best of our knowledge) been previously reported.
The emergence of the peak at $Z_i = 4$ coincides with the emergence of long high-$Z_i$ tails in the $P(Z_i)$ that correspond to large spheres contacted by many smaller spheres; related issues were discussed in Refs.\ \otherchanges{\cite{monti22,monti23,srivastava25}}.

In systems prepared \textit{with} SWAP and TDOF moves, the $P(Z_i)$ are radically different.
First, all floaters and rattlers with $Z_i \leq 2$ are eliminated.
The remaining rattlers all have $Z_i = 3$, and the $P(3)$ are \changedforRone{at least three} times lower than in systems prepared without these moves for all \changedforRone{$\Delta > 0$}.
The $P(4)$ are \changedforRone{lower (higher) for all $\Delta < 0.25$ ($\Delta \geq 0.25$).}
Finally, for all $Z_i > 4$, the $P(Z_i)$ are larger and decrease slower with increasing $Z_i$ \otherchanges{than in systems prepared without these moves}, particularly for $\Delta \gtrsim 0.2$.

All of the \otherchanges{protocol-dependencies illustrated in Figs.\ \ref{fig:2}-\ref{fig:oldfig2b}} can be explained as follows. 
LS-like particle-growth algorithms, \changedforRthree{when performed without SWAP moves,} terminate with  large numbers of floaters/rattlers occupying the gaps between the particles forming the jammed backbone.  
\changedforRthree{For $\Delta \lesssim 0.25$, adding SWAP moves to the growth process eliminates most of these rattlers.
For larger $\Delta$, SWAP moves are insufficient because many smaller particles remain free to rattle within cages formed by larger particles, but adding TDOF moves to the growth process allows these rattlers} to continue growing until most of them also jam.
We believe that the remaining rattlers indicate that the algorithm described in Sec.\ \ref{sec:methods} is imperfect.
Improving it might produce ideally-stable, rattler-free packings \otherchanges{-- analogous to those obtained in Ref.\ \cite{bolton24} --} \changedforRthree{for a wider range of $\Delta$.}

Relatively few studies of jamming have examined \textit{hyperstatic} packings, i.e.\ systems with $Z_{\rm J} > Z_{\rm iso} \equiv 6$.
Brito \textit{et al.}\  found that SWAP dynamics increase the number of contacts required to mechanically stabilize polydisperse packings, with $\delta Z = Z_{\rm J} - Z_{\rm iso} \sim \Delta^{1/2}$ \cite{brito18b}.
\changedforRone{Here we found that the combination of SWAP and TDOF moves produces substantailly-hyperstatic packings for all $\Delta > 0$.}
\otherchanges{The ($\phi_{\rm J},\Psi$)-related concepts discussed in Refs.\ \cite{torquato00,donev04c,donev05c,brito18b,anzivino23} indicate that such hyperstatic packings  must possess enhanced structural order, but they do not specify the form such order will take.
Moreover,} the connections between the hyperstaticity and mechanical stability of \otherchanges{our higher}-$\Delta$ packings are far less clear than is the case for their low-$\Delta$ counterparts.
For example, we find that the fraction of particles with $Z_i > 12$, which is impossible for monodisperse hard spheres, is nonzero for all \changedforRone{$\Delta \geq 0.10$.
This fraction} exceeds 1\% for all $\Delta > \changedforRone{0.26}$ ($\Delta > \changedforRone{0.19}$) in systems prepared without (with) SWAP and TDOF moves.
The emergence of long tails in the $P(Z_i)$ distributions for $\Delta \gtrsim 0.25$ indicates that caution must be exercised when using $Z_{\rm J}$ as a metric to assess high-$\Delta$ packings' mechanical stability.
\otherchanges{In the following sections, we will explore these questions in detail.}

\begin{figure}[h]
\includegraphics[width=3in]{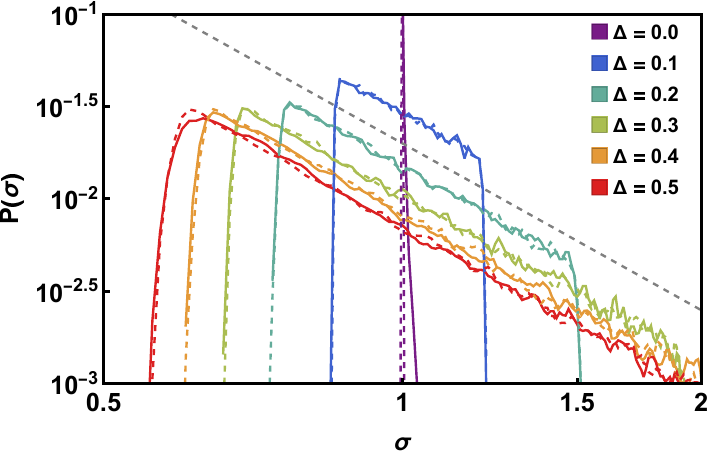}
\caption{Protocol dependence of particle-diameter distributions for $N = 1000$.  All lengths have been scaled to yield $\langle \sigma \rangle = 1$.  Solid and dashed curves show results obtained with and without SWAP and TDOF moves, and the dashed line indicates $P(\sigma) \sim \sigma^{-3}$ scaling.}
\label{fig:3}
\end{figure}

\subsection{Other structural-order metrics}
\label{subsec:IIIB}

One might expect that the changes in $Z_{\rm J}(\Delta)$ and $P(Z_i)$ enabled by employing TDOF moves would necessarily coincide with \otherchanges{substantial} changes in the jammed states' $P(\sigma)$ \changedforRthree{for $\Delta \gtrsim 0.25$.}
The latter would cast doubt on the utility of our results since TDOF moves are unphysical.
\otherchanges{However, a}s illustrated in Figure \ref{fig:3}, the changes in $P(\sigma)$ associated with employing TDOF moves are negligible for \otherchanges{all} $0 < \Delta \lesssim 0.45$.
For $\Delta = 0$, when the moves are not employed, $P(\sigma) = \delta(\sigma - 1)$.
When they are, $P(\sigma)$ is nonzero over a small range $\changedforRtwo{0.99} < \sigma < 1.04$; this is only a very small broadening.
For $\Delta \gtrsim 0.45$, $P(\sigma)$ decreases slightly for $\sigma \simeq \sigma_{\rm min}$ because the smallest particles are most able to grow as they fill in the jammed backbone's gaps.
We believe that this effect is responsible for the increase of $\phi_{\rm J}(\Delta)$ above its linear-fit value (Fig.\ \ref{fig:1}).

The TDOF moves employed in Refs.\ \cite{hagh22,bolton24} produced much larger changes in $P(\sigma)$ because they were applied \textit{throughout}  the particle-growth process. 
In contrast, our approach \otherchanges{(like that of Ref.\ \cite{kim24})} minimizes changes in $P(\sigma)$ by not implementing TDOF moves until near the end of this process.
Thus one can imagine that the effects of including TDOF moves highlighted in Fig.\ \ref{fig:2} could be at least partially reproduced in experiments, using a suitably designed packing-preparation protocol.

\begin{figure}[htbp]
\includegraphics[width=3in]{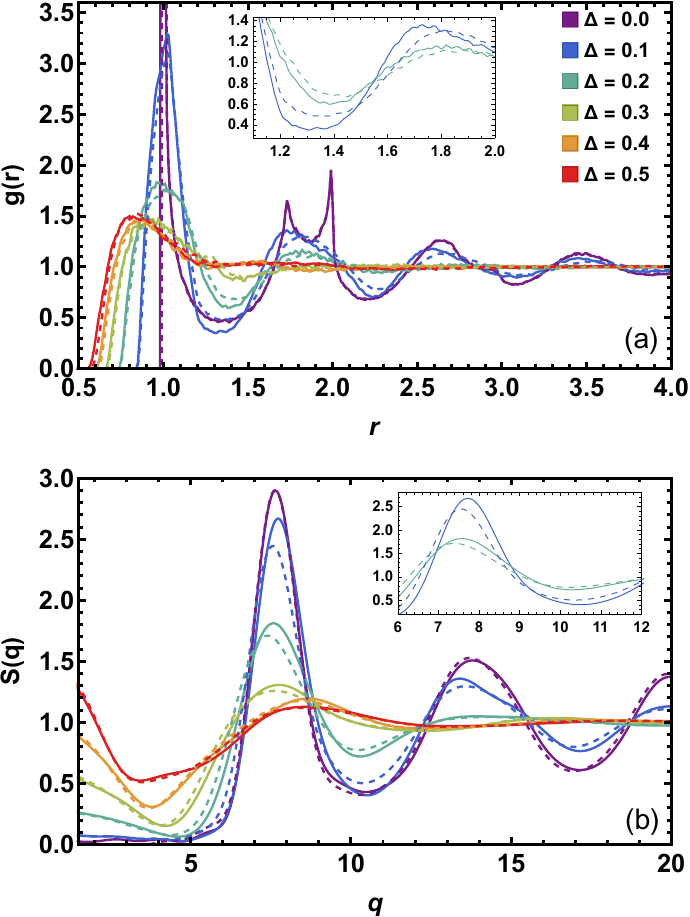}
\caption{Protocol dependence of pair correlation functions \changedforRtwo{and structure factors} for $N = 1000$.  
Solid and dashed curves show results obtained with and without SWAP and TDOF moves, and the  \changedforRtwo{insets highlight} the 
 \changedforRtwo{protocol-dependencies} for $\Delta = 0.10$ and $0.20$.  All lengths have been scaled to yield $\langle \sigma \rangle = 1$.  \otherchanges{Similar $\Delta$-dependence of $g(r)$ in packings prepared without SWAP was reported in Ref.\ \cite{oquendo20}.}}
\label{fig:5}
\end{figure}

\begin{figure*}
\includegraphics[width=6.65in]{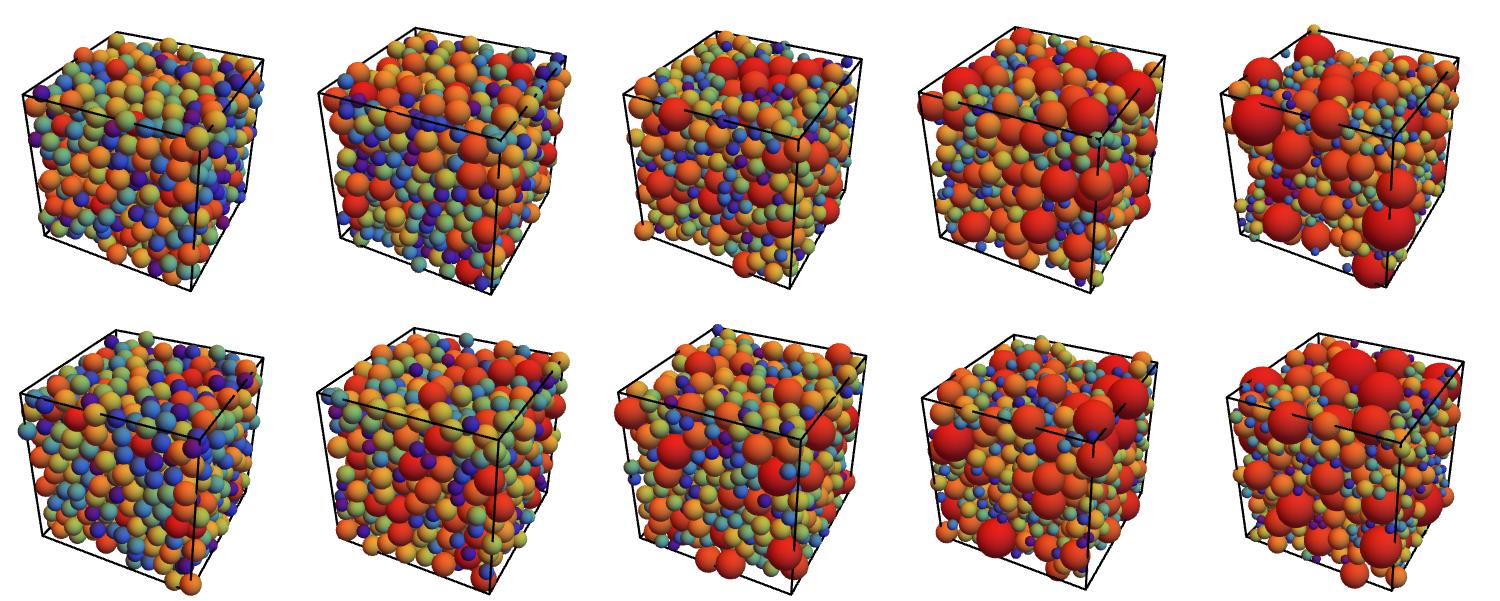}
\caption{Typical jammed states for $N = 1000$.  The top and bottom rows respectively show packings generated without and with SWAP and TDOF moves.  Columns from left to right show $\Delta = 0.10,\ 0.20,\ 0.30,\ 0.40,\ \rm{and}\ 0.50$.  Particle colors range from purple to red in order of increasing $\sigma$.   Only spheres whose centers lie within the boxes are shown.}
\label{fig:4}
\end{figure*}

Previous studies of how preparation protocol affects jammed packings' ($\phi_{\rm J},\Psi$) \cite{donev04c,ozawa17} have shown that for fixed $P(\sigma)$ and $\Delta$, systems with larger $\phi_{\rm J}$ have higher rattler fractions $f_{\rm ratt}$, more bond-orientational and icosahedral-like order, reduced Voronoi-cell asphericity, and larger low-$k$ spectral densities (i.e. smaller long-wavelength packing-fraction fluctuations \cite{zachary11b}).
We have already shown that employing TDOF moves reverses the $\phi_{\rm J}$-dependence of $f_{\rm ratt}$, and we do not attempt to duplicate the other structural analyses here.
Instead we ask a simpler question:\ does employing SWAP and TDOF moves during particle growth \textit{qualitatively} change packings' structural order?

Examining the packings' pair correlation functions $g(r)$ yields additional insights.
As illustrated in Figure \ref{fig:5}\changedforRtwo{(a)}, employing SWAP and TDOF moves can produce packings with better-organized coordination shells.
In particular, it yields a clearer separation between nearest and second-nearest neighbors, as indicated by the deeper first minima and higher second maxima of $g(r)$.
These differences, however, \textit{only occur for $\Delta \lesssim \otherchanges{0.25}$.}
For $\Delta \gtrsim \otherchanges{0.25}$, the $g(r)$ are nearly protocol-independent.
The increasingly-wide $P(\sigma)$ distributions (Fig.\ \ref{fig:3}) make the first and second coordination shells indistinguishable, and the $g(r)$ are nearly ideal-gas-like for $r > \sigma_{\rm max}$.

\changedforRtwo{Another useful metric for characterizing these packings' order is the static structure factor $S(q)$.
Fig.\ \ref{fig:5}\changedforRtwo{(b)} shows that the main differences in $S(q)$ produced by employing SWAP and TDOF moves directly reflect the abovementioned differences in $g(r)$.
More specifically, employing these moves increases the height of the first peaks in $S(q)$ and shifts them to larger $q$, \textit{but only for} $\Delta \lesssim 0.2$.
These trends directly correspond to the abovementioned deepening and shift to smaller $r$ of the first minima in $g(r)$.
Another result apparent from this figure is that while fractionation  increases with increasing $\Delta$ [as indicated by the increasing $\lim_{q\to 0} S(q)$], \textit{the degree to which this is so is protocol-independent.}
In other words, employing SWAP and TDOF moves does not seem to increase fractionation, as it did for hard-sphere liquids with $P(\sigma) \sim \sigma^{-3}$ and $\Delta \lesssim 0.10$ that were equilibrated at sufficiently large $\phi_{\rm eq}$ \cite{ninarello17}.}

\changedforRtwo{Are these structural differences} apparent to the naked eye?
Figure \ref{fig:4} shows that the changes are, in fact, rather subtle.
In particular, employing SWAP moves does \textit{not} produce fractionated \textit{polycrystalline} packings, as it did for low-$\alpha$ ellipses \cite{hoy24b}.
The only difference that is clear in these snapshots is that employing SWAP and TDOF moves allows small particles to more efficiently fill the gaps between the larger ones.
\changedforRtwo{Note, however, that we cannot rule out the presence of some local crystallization for small $\Delta$ near the peak in $Z_{\rm J}$  illustrated in Fig.\ \ref{fig:2}.}

\subsection{Bulk moduli}
\label{subsec:IIIB}

\begin{figure}[htbp]
\includegraphics[width=2.95in]{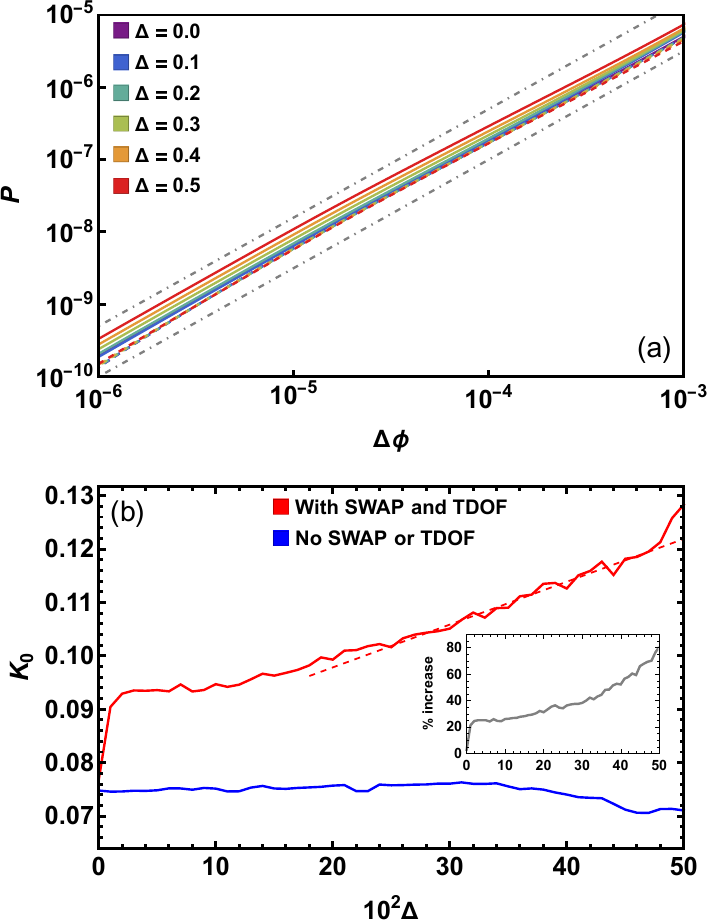}
\caption{Protocol dependence of \changedforRtwo{bulk moduli} for $N = 1000$.  
\changedforRtwo{Panel (a) shows results for $P(\Delta \phi)$, while panel (b) shows our estimates for $K_0$.
In panel (a), solid and dashed curves show results obtained with and without SWAP and TDOF moves, and the dot-dashed lines indicate $P \sim (\Delta \phi)^{3/2}$ scaling.}  
In panel (b), rhe dashed line shows a linear fit to $K_0$ over the range $0.25 \leq \Delta \leq 0.45$, and the inset shows the fractional increase in $K_0$ obtained by employing SWAP and TDOF moves.
 All lengths have been scaled to yield $\langle \sigma \rangle = 1$, and $K_0$ values are given in urits of $\varepsilon/\langle \sigma \rangle^3$. }
 \label{fig:6}
\end{figure}

Finally, to demonstrate that the various structural differences discussed above strongly affect the packings' mechanical properties, we examine their bulk moduli $K(\Delta)$.
Multiple previous studies have examined how $K$ depends on the pair potential, particle size ratio (in bidisperse systems), and the overcompression $\Delta \phi = \phi - \phi_{\rm J}$ \cite{ohern03,goodrich16,petit22,mizuno16}.
Here we focus on how $K$ depends on $\Delta$ and preparation protocol.
For Hertzian interactions, $P \sim (\Delta \phi)^{3/2}$ in the small-$\Delta \phi$, elastic regime where systems deform affinely \cite{ohern03}.
The resulting bulk moduli are singular at jamming, i.e.\ $K \sim (\Delta \phi)^{1/2}$, but a useful measure of their relative magnitudes can be obtained by writing 
\begin{equation}
K = \phi \displaystyle\frac{\partial P}{\partial \phi} \bigg{ | }_{\phi \simeq \phi_{\rm J}} \simeq \displaystyle\frac{3\phi_{\rm J} P_0}{2}  (\phi - \phi_{\rm J})^{1/2} \equiv K_0  (\phi - \phi_{\rm J})^{1/2}  .
\label{eq:K0vals}
\end{equation}
Figure \ref{fig:6} shows the $K_0$ values obtained by fitting the $P(\Delta \phi)$ data to $P = P_0 (\Delta \phi)^{3/2}$ over the power-law scaling regime, i.e.\ over the range \changedforRtwo{$10^{-6} \leq \Delta \phi \leq 10^{-3}$}, and then plugging the obtained $P_0$ values and results for $\phi_{\rm J}$ discussed above into Eq.\ \ref{eq:K0vals}.

In systems prepared without SWAP and TDOF moves, $K_0$ is nearly $\Delta$-independent for $\Delta \lesssim 0.35$, then decreases slowly with increasing $\Delta$ over the range $0.35 \lesssim \Delta \lesssim 0.50$.
We explain these trends as follows:
Bulk moduli can be written formally as $K = K_{\rm aff} - K_{\rm na}$, where
\begin{equation}
K_{\rm aff} = \displaystyle\frac{1}{V} \displaystyle\sum_{i = 1}^{N-1} \displaystyle\sum_{j = i+1}^N \left[ \displaystyle\frac{\partial^2 U}{\partial r_{ij}^2} - \displaystyle\frac{1}{r_{ij}} \displaystyle\frac{\partial U}{\partial r_{ij}}  \right] \displaystyle\frac{r_{ij}^2}{9}
\label{eq:Kaff}
\end{equation}
is the classical Born term arising from affine compression and $K_{\rm na}$ captures the reduction of $K$ by nonaffine particle displacements \changedforRone{\cite{maloney04,zaccone11,mizuno16,milkus16}}.
The $1/V$ prefactor and the purely-repulsive interactions [i.e., $U(r_{ij}) = 0$ for $r_{ij} > \sigma_{ij}$] respectively imply $K \sim \phi_{\rm J}$ and $K \sim Z_{\rm J}$ for \textit{monodisperse} systems.
Polydisperse systems are more complicated, as the $Z_{\rm J}$-dependence of their bulk moduli depends sensitively on which particles are in contact.
Our results suggest that as $\Delta$ increases over the range $0 \leq \Delta \lesssim 0.35$, the effects of increasing $\phi_{\rm J}$ (Fig.\ \ref{fig:1}), decreasing $Z_{\rm J}$ (Fig.\ \ref{fig:2}), and the fact that the sum in Eq.\ \ref{eq:Kaff} is increasingly dominated by the energetically-costly contacts between large particles that dominate the jammed backbone \cite{monti22,monti23} may roughly cancel each other out.

In systems prepared \textit{with} SWAP and TDOF moves, the $\Delta$-dependence of $K_0$ is dramatically different.
Instead of remaining nearly $\Delta$-independent over a wide range,  $K_0$ increases rapidly over the range $0 \leq \Delta \lesssim \changedforRtwo{0.03}$ before plateauing over the range \changedforRtwo{$0.03 \lesssim \Delta \lesssim 0.12$}.
Next $\partial K_0/\partial \Delta$ increases (with a negative $\partial^2 K_0/\partial \Delta^2$) over the range \changedforRtwo{$0.12 \lesssim \Delta \lesssim 0.25$}.
For $\changedforRtwo{0.25} \lesssim \Delta \lesssim 0.45$, $K_0$ increases linearly with $\Delta$ as indicated by the dashed line.
Finally, for $\Delta \gtrsim 0.45$, $K_0$ increases above this linear trend.
Clearly these trends in $K_0(\Delta)$ closely track the trends in $\phi_{\rm J}(\Delta)$ shown in Fig.\ \ref{fig:1}, especially for $\Delta \gtrsim \changedforRtwo{0.25}$.

\changedforRone{We believe that the higher $\phi_{\rm J}(\Delta)$ (Fig.\ \ref{fig:1}), larger $Z_{\rm J}(\Delta)$ (Fig.\ \ref{fig:2}) and better-organized coordination shells (Fig.\ \ref{fig:5}) combine to make} 
the packings generated using SWAP and TDOF moves far more mechanically-stable.
As shown in Fig.\ \ref{fig:6}'s inset, the fractional increases in $K_0(\Delta)$ obtained by employing these moves increase rapidly from \changedforRtwo{$\simeq 3\%$ to $\simeq 25\%$} over the range $0 \leq \Delta \lesssim \changedforRtwo{0.03}$, then increase more gradually (from \changedforRtwo{$\simeq 25\%$ to $\simeq 45\%$}) over the range $\changedforRtwo{0.03} \leq \Delta \lesssim 0.35$.
For larger $\Delta$, the decreasing $K_0$ for systems prepared without these moves and increasing $\partial K_0/\partial \Delta$ for systems prepared with these moves produce even-more-dramatic fractional increases in $K_0$, reaching a maximum of \changedforRtwo{80\%} for $\Delta = 0.50$.

\changedforRone{Here, since we are focusing mainly on structure rather than mechanics, we presented only a brief analysis of systems' bulk moduli, to illustrate the basic result that polydisperse 3D sphere packings generated using SWAP and TDOF moves are far more mechanically-stable than packings generated without these moves, and that the magnitude of these stability increases grows rapidly with increasing $\Delta$.
We have chosen not to speculate about the role nonaffine deformation plays in the above results because rigorously testing any hypotheses about $K_{\rm na}$ would require detailed analyses  \cite{maloney04,zaccone11,mizuno16,milkus16} that are beyond our present scope.
We plan to explore this issue further in a followup study that wlll characterize these packings' linear and nonlinear mechanics, under both compression and shear, in far more detail.}

\section{Discussion and Conclusions}
\label{sec:conclude}

Many real granular materials are highly polydisperse.  
For example, power-law-like particle-size distributions $P(\sigma) \sim \sigma^{-\beta}$ over two or more orders of magnitude in $\sigma$ are found in natural systems ranging from soils \cite{wu93} to sea ice \cite{rothrock84}, and in man-made materials produced by a wide range of comminution/fragmentation processes \cite{turcotte86}. 
Simulations can systematically study the effects of this dispersity by employing variable-width $P_{\otherchanges{\mathcal{R}}}(\sigma)$ with $\mathcal{R}$-dependent $\Delta$, e.g.\ those given by Eqs.\ \ref{eq:Pofsigma}-\ref{eq:deltaform}, but suprisingly few studies have done so.
In particular, the preparation-protocol-dependence of amorphous solids' structure and mechanics remains a very active field of research with many open questions \cite{jin18,richard20,berthier25}, but most theoretical studies of these questions have employed models with a single, fixed particle dispersity $\Delta \lesssim 0.25$.

In this paper, we have shown that the dramatic increases in jammed sphere packings' ($\phi_{\rm J},\Psi$) achievable by thermally equilibrating them at a higher packing fraction $\phi_{\rm eq}$ \cite{chaudhuri10,morse21,berthier16b,ozawa17}, compressing them slower \cite{torquato00}, artificially speeding up their  dynamics during sample compression \cite{ghimenti24,berthier24}, or introducing additional degrees of freedom associated with the particle diameters \cite{kapteijns19,hagh22,bolton24} can be achieved for a very wide range of particle dispersities ($0 < \Delta \leq 0.50$), but the effects of employing such techniques can vary strongly and nontrivially with $\Delta$.
\changedforRtwo{The $\Psi$ investigated here included $Z_{\rm J}$, $P(Z_i)$, $g(r)$ for intermediate $r$, and $S(q)$ for intermediate $q$; we anticipate that comparable effects will be observable for other $\Psi$.}

We showed that the packing-efficiency/mechanical stability gains obtained by employing SWAP and TDOF moves increase \otherchanges{almost} monotonically with $\Delta$.  
This result was surprising because the utility of standard SWAP for equilibrating deeply supercooled liquids decreases for $\Delta \gtrsim 0.25$ \cite{ninarello17}; we believe that it holds because our algorithm still yields $\gg 1$ successful swaps per particle over the course of the particle-growth process.
As summarized in Section \ref{sec:results}, packing-efficiency/mechanical stability gains tend to increase rapidly over the range $0 < \Delta \lesssim 0.1$, then slowly over the range $0.1 < \Delta \lesssim 0.25$, then linearly over the range $0.25 < \Delta \lesssim 0.45$, then more-rapidly for \otherchanges{$\Delta \gtrsim 0.45$} as the effects of including TDOF moves at the end of the process grow stronger.

Including the latter moves offers several advantages over algorithms that only employ SWAP.
For example, they allow one to achieve both  higher $\phi_{\rm J}$ \textit{and} lower $f_{\rm ratt}$ \otherchanges{for $\Delta \gtrsim 0.25$ \cite{hagh22,bolton24,kim24}}, whereas standard algorithms yield increasing $f_{\rm ratt}(\phi_{\rm J})$ \cite{ozawa17,monti22,monti23}.
This combination \otherchanges{has been shown} to produce packings that are unusually mechanically/vibrationally stable at $T = 0$, and have ideal-glass-like properties at finite $T$ \cite{bolton24}.
We have shown that \otherchanges{it} can \changedforRtwo{nearly} \textit{double} the packings' bulk moduli for $\Delta \gtrsim 0.45$, and attributed these large gains to the fact that TDOF moves allow small rattlers that are \textit{not} initially part of the packings' mechanically-rigid backbones to continue growing \otherchanges{until} they, too, jam.
Even-more-dramatic stability gains are expected for the packings' shear moduli $G(\Delta)$ because \changedforRone{the} suppression of nonaffine deformation \changedforRone{associated with the increased $Z_{\rm J}$ should} increase $G$ much more than it increases $K$ \changedforRone{\cite{ohern03,zaccone11}}.

Here we only studied systems with $\beta = 3$; this value  was chosen primarily because it has been employed in multiple recent studies of the glass-jamming transition \otherchanges{\cite{interiano24,ozawa17,berthier24,berthier16,ninarello17,scalliet22,oquendo20,oquendo22,monti22,monti23,anzivino23,srivastava25}}.
We anticipate, however, that the qualitative trends discussed above will also be present for other values of $\beta$.  
$\beta = 3.5$ should be an especially-interesting case for further study, as it yielded the highest $\phi_{\rm J}(\Delta)$ in Refs. \otherchanges{\cite{oquendo20,oquendo22,monti22,monti23}}, and also corresponds to the $P(\sigma)$ of grains generated by explosive rock fragmentation \cite{turcotte86}.
We also anticipate that at least some of these trends will be present for other short-ranged purely-repulsive interaction potentials $U(r)$.  
If so, this will present multiple opportunities to answer open questions, because the differences between ultrastable/ideal glasses and their less-stable counterparts  are comparable to the differences between high-($\phi_{\rm J},\Psi$) jammed states and their lower-($\phi_{\rm J},\Psi$) counterparts \cite{swallen07,ediger17,kapteijns19,hagh22,bolton24,kim24,berthier25}.
Exceptionally-stable soft-sphere glasses \otherchanges{have been prepared recently} using thermal cSWAP \cite{nishikawa25}.
\otherchanges{Adding TDOF moves to the end of such sample-preparation procedures could make them even more stable.}

\otherchanges{We thank Joseph Monti and Alessio Zaccone for helpful discussions.}
This material is based upon work supported by the National Science Foundation under Grant Nos. DMR-2026271 and DMR-2419261.

\begin{appendix}

{\color{purple}
\section{Comparison to Eq. 5 and its associated theory}

Anzivino \textit{et al.}\ obtained both  Eq.\ \ref{eq:zacconephiJ} and analytic expressions for its parameters  $\{a_1,a_2,a_3,b_1,b_2,b_3\}$ \cite{anzivino23}:
\begin{equation}
\begin{array}{c}
a_1 = \displaystyle\frac{5 - (8C_0 + 5)\phi_{\rm mono}^{\rm RCP} }{2C_0 (1 -  \phi_{\rm mono}^{\rm RCP})Z'(\phi_{\rm mono}^{\rm RCP})} \\
\\
a_2 = \displaystyle\frac{6[1 -  (C_0+1) \phi_{\rm mono}^{\rm RCP}]}{C_0 (1 -  \phi_{\rm mono}^{\rm RCP})Z'(\phi_{\rm mono}^{\rm RCP})} \\
\\
a_3 = \displaystyle\frac{(4C_0+1) \phi_{\rm mono}^{\rm RCP}}{2C_0 (1 -  \phi_{\rm mono}^{\rm RCP})Z'(\phi_{\rm mono}^{\rm RCP})} \\
\\
b_1 = \displaystyle\frac{7}{2} + \displaystyle\frac{4}{(1 -  \phi_{\rm mono}^{\rm RCP})^2 Z'(\phi_{\rm mono}^{\rm RCP})} \\
\\
b_2 = 3 + \displaystyle\frac{6}{(1 -  \phi_{\rm mono}^{\rm RCP})^2 Z'(\phi_{\rm mono}^{\rm RCP})} \\
\\
b_3 = \displaystyle\frac{1}{2} - \displaystyle\frac{2}{(1 -  \phi_{\rm mono}^{\rm RCP})^2 Z'(\phi_{\rm mono}^{\rm RCP})} \\
\end{array}.
\label{eq:zaccparams}
\end{equation}
Here $\phi_{\rm mono}^{\rm RCP} \equiv \phi_{\rm J}(0)$, $\mathcal{Z} =\frac{P}{\rho k_B T}$ is the compressibility factor, and $\mathcal{Z}'(x) = \frac{d\mathcal{Z}}{d\phi} |_x$ can be evaluated using a liquid-state-theoretic equation of state for $\mathcal{Z}(\phi)$.
We choose to employ the Carnahan-Starling expression \cite{carnahan69}
\begin{equation}
\mathcal{Z}_{\rm CS}(\phi) =  \displaystyle\frac{1}{1-\phi} +  \displaystyle\frac{3\phi}{(1-\phi)^2} +  \displaystyle\frac{\phi^2 (3-\phi)}{(1-\phi)^3}.
\label{eq:CSEOS}
\end{equation}
Thus Eq.\ \ref{eq:zacconephiJ} is a function only of $\phi_{\rm mono}^{\rm RCP}$, $C_0$ and $\Delta$.

\begin{figure}[htbp]
\includegraphics[width=3in]{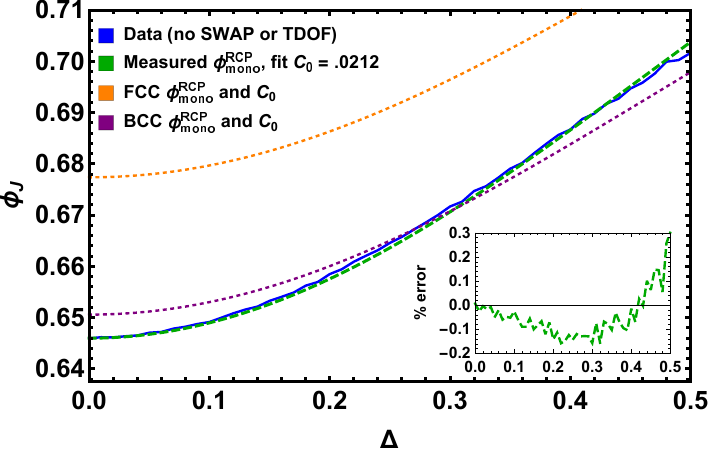}
\caption{\changedformultiple{Comparison of jamming densities for $N = 1000$ to the theory of Anzivino \textit{et al.}
The dotted orange and purple curves respectively show the predictions of Eq.\ \ \ref{eq:zacconephiJ} for fcc and bcc boundary conditions  \cite{anzivino23}, i.e.\ for  ($\phi_{\rm mono}^{\rm RCP} = 0.6774$, $C_0 = \changedforRtwo{0}.01874$) and ($\phi_{\rm mono}^{\rm RCP} = 0.6506$, $C_0 = \changedforRtwo{0}.0246$). The green dashed curve show the predictions of Eq.\ \ \ref{eq:zacconephiJ} for our measured $\phi_{\rm mono}^{\rm RCP} = 0.646$ and fit $C_0 = .02118$, and  inset shows the fractional error of this prediction relative to our simulation data for $\phi_{\rm J}(\Delta)$.}}
\label{fig:7}
\end{figure}

Anzivino \textit{et al.}\ chose to calculate both $C_0$ and $\phi_{\rm mono}^{\rm RCP}$ from first principles.
Their calculation required assuming  that the liquid will form either a fcc crystal or a bcc crystal at sufficiently large $\phi$; these two different ``boundary conditions'' yielded different values of $\phi_{\rm mono}^{\rm RCP}$ and $C_0$ \cite{anzivino23}.
Figure \ref{fig:7} compares the predictions of their theory to our data for $\phi_{\rm J}(\Delta)$ in systems prepared without SWAP or TDOF moves.
Both predictions qualitatively capture the observed trends, but exhibit substantial quantitative disagreement with our simulation data; Ref.\ \cite{anzivino23} reported similar quantitative disagreements.

An alternative approach is to take the simulated $\phi_{\rm mono}^{\rm RCP} \equiv \phi_{\rm J}(0)$ as a known input, and then fit the data for $\phi_{\rm J}(\Delta)$ to Eq.\ \ref{eq:zacconephiJ} with $C_0$ as the only fitting parameter.
Using this approach, we obtain $C_0 \simeq 0.0212$.
As shown in Fig.\ \ref{fig:7}, this value predicts the simulated $\phi_{\rm J}(\Delta)$ to within less than 0.3\% over the entire range $0 \leq \Delta \leq 0.5$.
This agreement is remarkable given that Eq.\ \ref{eq:zaccparams} takes only quantities associated with monodisperse systems [i.e., $\phi_{\rm mono}^{\rm RCP}$ and $Z'(\phi_{\rm mono}^{\rm RCP})$] as its inputs and Eq.\ \ref{eq:CSEOS} is an equation of state for monodisperse spheres.

On the other hand, Eq.\  \ref{eq:zacconephiJ} clearly does \textit{not} accurately predict even the shape of the $\phi_{\rm J}(\Delta)$ curve for systems prepared using SWAP and TDOF moves.
A similar issue arises for ellipses, where a physically-motivated analytic functional form (Eq.\ 13 of Ref.\ \cite{rocks23}) predicts the $\phi_{\rm J}(\alpha)$ of ellipse packings to within less than 0.3\% over the range $1 \leq \alpha \leq 10$ for systems prepared without these moves, but fails to capture the shape of $\phi_{\rm J}(\alpha)$ for systems prepared with these moves \cite{hoy24b}.
These results indicate that the preparation-protocol-dependence of $\phi_{\rm J}$ couples to $\Delta$ and $\alpha$ in a way that is not accounted for by existing theories.
Resolving this issue could be a critical contribution to the ongoing effort to analytically predict the structural and mechanical properties of jammed packings \changedforRone{\cite{zaccone22,zaccone25,zaccone25b}}.\newline
}

{\color{purple}
\section{Rattler-related considerations}

In soft-sphere packings, contacts are unambiguously defined as interparticle overlaps: $Z_i = \sum_{j \neq i} \Theta(\sigma_{ij} - r_{ij})$ where $\Theta$ is the Heaviside step function, and the average coordination number $Z = N^{-1} \sum_{i = 1}^N Z_i$.
For hard-sphere packings, however, interparticle overlaps are non-existent, and the above definition of $Z_i$ gives $Z = 0$.
Many authors have instead employed the $C$-dependent contact criterion discussed in Section \ref{subsec:IIIA}, i.e.\ $Z_i = \sum_{j \neq i} \Theta[(1+C)\sigma_{ij} - r_{ij}]$, but this approach gives strongly-$C$-dependent $Z$ for small $C$ owing to the divergence of the scaled pair correlation function $g(r_{ij}/\sigma_{ij})$ as $r_{ij}$ approaches $\sigma_{ij}$ from above \cite{ohern03,donev05c,ozawa17,anzivino23}.

One can mitigate this issue by choosing $C$ values that lie within the well-known plateau in $Z(C)$ \cite{donev05c}, but this complicates rattler identification.
For hard-sphere packings, most previous studies have identified rattlers as particles with $Z_i < 4$ \cite{noncohem}.
Many well-known results for marginally-jammed packings, e.g.\ the characteristic isostaticity criterion $Z_{\rm J} = 6$ and scaling laws such as $Z(\phi) - Z_{\rm J} \sim \sqrt{\phi - \phi_{\rm J}}$, were obtained in systems where rattlers were \textit{recursively} removed prior to calculating $Z$ \cite{ohern03,zaccone11,ozawa17,goodrich16}.
Unfortunately, as illustrated in Figure \ref{fig:9}, this procedure couples to the $C$-dependence of $Z$.

\begin{figure}[htbp]
\includegraphics[width=3in]{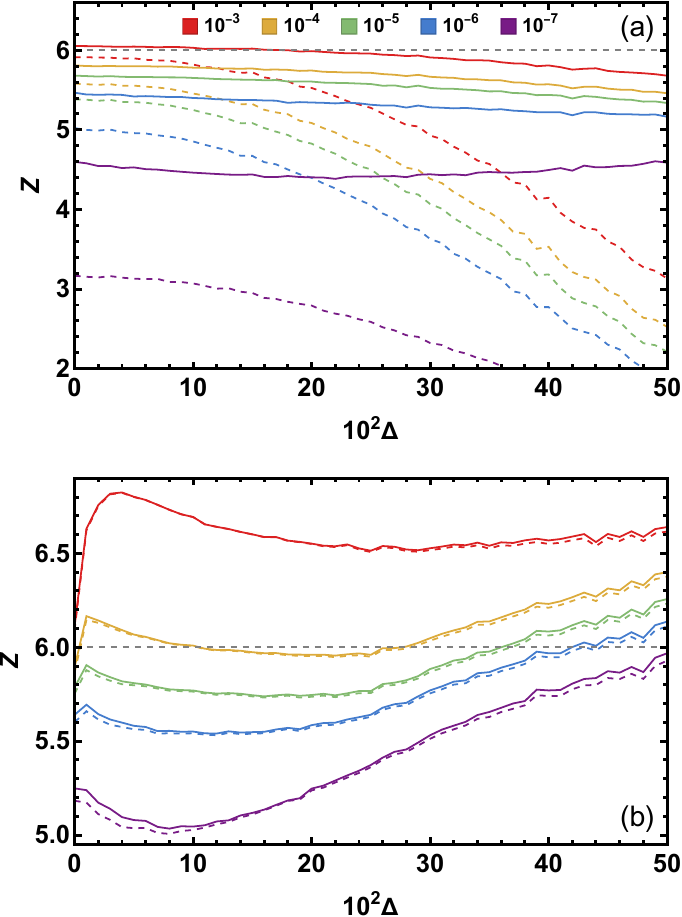}
\caption{{\color{purple} Protocol dependence of average coordination numbers and rattler fractions for $N = 1000$ and various values of $C$ indicated by the color legend.  Panel a (b) shows results for systems prepared without (with) SWAP and TDOF moves.  Solid (dashed) curves show results for the average coordination number calculated after (before) recursively removing particles with $Z_i < 4$.}}
\label{fig:9}
\end{figure}

Consistent with the notion that employing SWAP moves allows the $\Gamma \to 0$ limit to be accessed, the differences between the results shown in panels (a-b) are analogous to the differences between results for high and low $\Gamma$ discussed in Ref.\ \cite{donev05c}.
Here we chose $C = 10^{-3}$ to perform the analyses described in Section \ref{sec:results}  because it yields $Z_{\rm J}(\Delta) \simeq 6$ for all $\Delta \lesssim 0.25$ in systems prepared without SWAP and TDOF moves, after particles with $Z_i < 4$ are recursively removed.
This allows our results to be straightforwardly compared with the numerous previous studies that have found $Z_{\rm J} = 6$.
}

\end{appendix}

%

\end{document}